\documentclass{jfm}
\usepackage{graphicx}
\usepackage{epstopdf, epsfig}
\usepackage{amsmath}
\usepackage{color}

\shorttitle{On the role of the Prandtl number in convection driven by heat sources and sinks}
\shortauthor{B. Miquel, V. Bouillaut, S. Auma\^itre and B. Gallet}
\graphicspath{{./}}

\newcommand{\ez}{\mathbf{\hat{e}}_z}
\newcommand{\prandtl}{\Pran}
\newcommand{\heatcap}{C_p}
\newcommand{\deltaTbulk}{\Delta T_\mathrm{bulk}}
\newcommand{\deltaTBL}{\Delta T_\mathrm{BL}}
\newcommand{\reynolds}{\Rey}
\newcommand{\rayleighQ}{Ra_Q}
\newcommand{\rayleigh }{Ra}
\newcommand{\nusselt  }{Nu}
\newcommand{\rayexpo}{\gamma}
\newcommand{\pranexpo}{\chi}
\newcommand{\zequi}{z_*}
\newcommand{\dimlessell}{\widetilde{\ell}}

\newcommand{\cor}[1]{{#1}}
\newcommand{\modif}[1]{{#1}}
\newcommand{\com}[1]{{#1}}

\title{On the role of the Prandtl number in convection driven by heat sources and sinks}

\author{Benjamin Miquel\aff{1}
  \corresp{\email{benjamin.miquel@cea.fr}},
  Vincent Bouillaut\aff{1},
  S\'ebastien Auma\^itre\aff{1},
 \and Basile Gallet\aff{1}}

\affiliation{\aff{1}Universit\'e Paris-Saclay, CEA, CNRS, SPEC, 91191, Gif-sur-Yvette, France
}

\begin{document}

\maketitle

\begin{abstract}
We report on a numerical study of turbulent convection driven by a combination of internal heat sources and sinks. Motivated by a recent experimental realisation~\citep{lepotPNAS18}, we focus on the situation where the cooling is uniform, while the internal heating is localised near the bottom boundary, over approximately one tenth of the domain height. We obtain scaling laws $\nusselt \sim \rayleigh ^\rayexpo \prandtl^\pranexpo$ for the heat transfer as measured by the Nusselt number $\nusselt$ expressed as a function of the Rayleigh number $\rayleigh$ and the Prandtl number $\prandtl$. After confirming the experimental value $\rayexpo\approx1/2$ for the dependence on $\rayleigh$, we identify several regimes of dependence on $\prandtl$. For a stress-free bottom surface and within a range as broad as $\prandtl \in [0.003, 10]$, we observe the exponent $\chi\approx 1/2$, in agreement with Spiegel's mixing length theory. For a no-slip bottom surface we observe a transition from $\chi\approx 1/2$ for $\prandtl \leq 0.04$ to $\chi\approx 1/6$ for $\prandtl \geq 0.04$, in agreement with scaling predictions by Bouillaut et al. The latter scaling regime \com{stems from heat accumulation in the stagnant layer} adjacent to a no-slip bottom boundary, which we characterise by comparing the local contributions of diffusive and convective thermal fluxes.
\end{abstract}


\section{Introduction}
Thermal forcing is an important driver for turbulence in numerous natural and industrial flows. Buoyancy greatly impacts the oceanic dynamics of the Earth~\citep{sutherlandPRF19} and of ice-clad moons~\citep{miquelPRF18,soderlundGRL19}, atmospheric flows~\citep{yanoNATURE03}, and internal dynamics of planets and stars~\citep{aurnouPEPI12,guervillyNAT19}. Implications and consequences range from climate modelling~\citep{plantYano} to the generation of large scale magnetic field via the dynamo effect~\citep{browningApJ08,soderlundEPSL12,calkinsJFM15}, to cite a few. A global characterisation of such convective flows boils down to the critical issue of turbulent heat transport: How is the enhanced heat flux related to the temperature inhomogeneities\com{, the geometry of the system and the properties of the fluid}? Owing to the extremely turbulent nature of the natural and industrial flows of interest, one tries to answer the questions above by identifying scaling-laws that capture the essential mechanisms at play and are amenable to extrapolations. 

Here we study Convection driven by Internal Sources and Sinks (CISS), whose scaling behavior departs from the standard setups of thermal convection, such as the Rayleigh--B\'enard convection (RBC)~\citep{Ahlers, chilla_schumacher}, horizontal convection (HC)~\citep{vreugdenhilJFM17, rochaJFM20}, or even internally-heated convection (IHC) with a cooling top surface~\citep{goluskinJFM15} or between fixed-temperature plates~\citep{goluskinJFM16}. Indeed, the latter setups all display thermal boundary layers that greatly throttle the heat flux, whereas these boundary layers are bypassed in CISS, which results in a more efficient heat transport scaling regime. For instance, a recent experimental and numerical study of CISS  in water by~\citet{lepotPNAS18} investigated the dependence of the heat flux with the temperature difference in the fluid. The dimensionless counterpart to this question amounts to finding a relationship between the standard Nusselt number and Rayleigh number (see equation~\ref{eq:output_parameters} for definitions). \citet{lepotPNAS18} measured $\nusselt \sim \rayleigh ^\rayexpo$ with $\rayexpo \approx 0.5$. This is in strong contrast with RBC experiments and numerical studies where $0.27 \leq \rayexpo \leq 0.39$~\citep{castaingJFM89,chavannePRL97,niemelaNAT00, Ahlers,hePRL12,DoeringAbsence}. This range of measured exponents support the fact that diffusivity coefficients play a role in the RBC setup, through the formation of boundary layers that throttle the heat flux. 

The scope of the present \com{study} is to investigate the variation of the heat flux as the Prandtl number (defined as the ratio of the fluid diffusivity coefficients, see eq.~\ref{eq:control_parameters}) varies. Even for the much studied RBC case, such studies remain scarce. In the laboratory, low Prandtl numbers are reached by employing liquid metal~\citep{fauveJPL81,
	cioniJFM97, aurnouJFM18}, which severely hinders measurements and virtually forbids visualizations. Direct numerical simulations (DNS) exhibit much more flexibility from the standpoint of diffusivity coefficients and have proved invaluable for understanding the role on the Prandtl number for both RBC~\citep{shishkinaPRF17} and HC~\citep{shishkinaPRL16}. In the present \com{study}, we employ DNS to explore the parameter space \cor{of} CISS. In particular, \cor{we put to scrutiny} recent predictions by~\citet{bouillautJFM19} that not only provide a model for \com{the} experimental results of~\citet{lepotPNAS18}, but also predict two distinct regimes characterized by different \cor{scaling behaviours with respect to the Prandtl number}.

The paper is laid out as follows. The formulation of the governing equations and the numerical methods are presented in section~\ref{sec:methodology}. Models for the heat transfer law are presented in section~\ref{sec:scaling_arguments} and the existence of two different regimes is predicted, depending on the Prandtl number. These regimes are evidenced in a data set obtained from direct numerical simulations (DNS) and presented in section~\ref{sec:2regimes}, together with an analysis of the thermal boundary layers. 
Finally, section~\ref{sec:discussion} contains a discussion and closing remarks. 


\section{Governing equations and numerical procedure}\label{sec:methodology}
\subsection{Model and parameters}
We consider a horizontal fluid layer of depth $H$, \modif{kinematic} viscosity $\nu$, \com{thermal} diffusivity $\kappa$, \com{specific heat capacity} $\heatcap$, and \modif{thermal expansion coefficient} $\alpha$, all of which are considered constant. The layer is subjected to \com{the} vertical gravitational acceleration of \com{(uniform)} magnitude $g$. \com{We consider the Boussinesq approximation, where the density $\rho$ is constant and uniform, except in the expression of the buoyancy force}. Volumic heat sources and sinks are present in the bulk of the fluid. In the experiment by~\citet{lepotPNAS18} and~\citet{bouillautJFM19}, a powerful spotlight shines visible light through the transparent bottom of the tank, into dyed water where absorption occurs over a typical length $\ell$. Following Beer-Lambert's law, the intensity of light decreases exponentially \cor{with the height $Z$, measured from the bottom surface}. Thus, we model the experimental radiation-induced heating with an exponentially decreasing volumic heat-source in the present numerical work:
\begin{equation}
Q_L(Z) = \frac{P}{\ell}\frac{\exp\left(-Z/\ell\right)}{1-\exp\left(-H/\ell\right)},
\end{equation}
where $P$ is the incident light flux at the bottom of the tank (power per unit area), which gets deposited into the fluid through absorption. In the laboratory experiment, this internal heat source \modif{yields ``secular heating''} \com{of the fluid}, which amounts to \com{a quasi-stationary state with} homogeneous internal cooling $Q_c(Z) = P/H$ balancing the radiative heat input, so that the net power deposited in the fluid layer vanishes: $\int_0^{H} \left( Q_L(Z) - Q_c(Z) \right) \mathrm{d}Z = 0$. 
We stress the fact that, motivated by the experimental realisations, we are concerned with a distribution of heat sources that takes on finite values at the boundaries. In that respect, the problem of interest here crucially differs from previous studies by~\citet{barkerApJ14}, who considers a distribution of heat sources and sinks that vanishes at the boundaries.

The characteristics of the setup can be cast into three dimensionless control parameters, namely the Prandtl number $\prandtl$, the flux-based Rayleigh number $\rayleighQ$, and the dimensionless light \cor{absorption-length} $\dimlessell$:
\begin{equation}
\prandtl = \frac{\nu}{\kappa}\,;\quad \rayleighQ=\frac{\alpha g PH^4}{\rho \heatcap\kappa^2 \nu}\,; \quad \dimlessell = \frac{\ell}{H}\,.
\label{eq:control_parameters}
\end{equation}

The response of the fluid to the imposed flux may be characterised by the magnitude of the temperature \modif{difference} $\Delta T$ that \com{appears} in the flow, the characteristic velocity of which is $U$. To make contact with the numerous RBC literature, one can then define the Reynolds number $\reynolds$, the temperature-based Rayleigh number $\rayleigh$, and the Nusselt number $\nusselt$:
\begin{equation}
\reynolds = \frac{UH}{\nu}\, ; \quad \rayleigh = \frac{\alpha g \Delta T H^3}{\nu\kappa}\, ; \quad \nusselt = \frac{PH}{\rho \heatcap \kappa \Delta T}\, ; \quad \mathrm{with}\quad \rayleigh_Q = \rayleigh\nusselt\, .
\label{eq:output_parameters}
 \end{equation}
\subsection{Dimensionless equations and parameters}
The Boussinesq equations are written in a dimensionless form by counting lengths in units of the depth $H$, time in units of diffusive time $H^2/\kappa$, and temperature in units of $\nu\kappa / \alpha g H^3$. Thus, the governing equations for the solenoidal velocity field ($\boldsymbol{\nabla\cdot u}=0$) and temperature field $\theta$ are:
\begin{subequations}
\begin{gather}
\prandtl^{-1} \left(\partial_t \boldsymbol{u} + \boldsymbol{u\cdot \nabla u}\right) = - \nabla p + \theta\,\ez + \nabla^2 \boldsymbol{u}\, ,\label{eq:NS_momentum}\\
\partial_t \theta + \boldsymbol{u\cdot \nabla}\theta = \nabla^2 \theta + \rayleighQ \left(\frac{\exp\left(- z/\dimlessell\right)}{\dimlessell \left[1-\exp\left(-1/\dimlessell\right)\right]} - 1\right)\, ,\label{eq:NS_heat}
\end{gather}
\label{eqs:Navier_Stokes}
\end{subequations}
\cor{where $z=Z/H$.} These governing equations are supplemented with adiabatic boundary conditions for temperature: $\partial_z \theta =0$ at $z=0,\, 1$. Kinematic boundary conditions are always impenetrable ($w=0$) and either no-slip ($u=v=0$) or stress-free ($\partial_z u = \partial_z v = 0$) {at top and bottom}. \cor{Finally, we consider} periodic boundary conditions \cor{ in} the horizontal. 



\subsection{Numerical procedure}
Direct numerical simulations (DNS) of equations~(\ref{eqs:Navier_Stokes}) are performed with the HPC code \emph{Coral}, which employs a Chebyshev-Fourier-Fourier decomposition in space and a family of implicit-explicit timestepping schemes. The boundary conditions in the vertical direction $z$ are enforced using Galerkin basis recombination. Computations are initialised with either small-amplitude noise or a previous solution and are carried out until a statistically stationary regime is \cor{clearly identified}. \emph{Coral} was already used for studying CISS \com{in}~\citet{miquelPRF19}\com{, where it is benchmarked against analytical solutions}. \com{The non-periodic version of \emph{Coral} has also been benchmarked against the numerical solutions obtained by~\citet{lepotPNAS18} using the solver \emph{Dedalus}~\citep{burnsPRR20}, for $\prandtl=1$ and $7$.} 

\section{Scaling arguments}\label{sec:scaling_arguments}

\cor{For completeness, we recall in broad strokes the scalings predicted by~\citet{bouillautJFM19}. Experiments and numerical simulations indicate that the large-scale flow consists \com{of turbulent convection cells} of aspect ratio close to one, the velocity of which obeys the free-fall scaling-law $U \sim \sqrt{\alpha g \Delta T H}$.}
\subsection{Bulk transport}

\cor{As a fluid element circles around \com{a turbulent convection cell}, it rapidly heats up in the region $z \lesssim \dimlessell$, while it slowly cools down in the region $z \gtrsim \dimlessell$. In other words, the heat input inside the region $z \lesssim \dimlessell$ is carried outside the heating region by the flow, which amounts to a balance between the source term and the \com{horizontal} advection term in the heat equation:}
\begin{equation}
\rho \heatcap U \partial_x T \sim \frac{P}{\ell}\,. \label{eq:bulk_balance}
\end{equation}
Upon injecting the free-fall velocity estimate, the balance above yields \com{an estimate $\deltaTbulk$ for the temperature difference between the heating region $z \lesssim \dimlessell$ and the bulk region $z \gtrsim \dimlessell$}. In dimensional and dimensionless form:
\begin{equation}
\left( \deltaTbulk  \right) ^3 \sim \frac {P^2H}{\alpha g \rho^2 \heatcap^2 \ell^2}\,;\quad \nusselt \sim \dimlessell \sqrt{\rayleigh\prandtl} \label{eq:nusselt_RaPr_bulk}\,.
\end{equation}
\com{The reasoning above and the associated $\deltaTbulk$ do not take into account the velocity boundary conditions at the bottom plate. In the next section, we show that the latter can induce a possibly stronger temperature difference $\deltaTBL$ between the immediate vicinity of the bottom boundary and the heating region $z \lesssim \dimlessell$. The total temperature difference between the bottom plate and the bulk region $z \gtrsim \dimlessell$ is then estimated as $\Delta T=\deltaTBL+\deltaTbulk$.}

\subsection{\cor{Stagnant layers near no-slip boundaries}}

\cor{The scaling-law (\ref{eq:nusselt_RaPr_bulk}) may be modified in the case of a no-slip bottom boundary condition. Indeed, the stagnant layer in the immediate vicinity of the bottom plate accumulates heat, and without efficient advection outside this region, the heat can only be diffused away. This leads to a thermal boundary layer of thickness $\delta_{\Theta}$. Balancing the heat source term with the diffusive one estimated inside the boundary layer leads to:}
\begin{equation}
\rho \heatcap \kappa \frac{\deltaTBL}{\delta^2_\Theta} \sim \frac{P}{\ell}\,, \qquad \mathrm{yielding:} \qquad 
\deltaTBL = \frac{PH^2}{\rho \heatcap \kappa \ell }\frac{\delta_\Theta^2}{H^2}\,.
\label{eq:temperature_jump_BL}
\end{equation}


\cor{A final relation between the Nusselt number and the Rayleigh and Prandtl numbers is obtained by supplementing (\ref{eq:temperature_jump_BL}) with a scaling estimate of the boundary layer thickness $\delta_\Theta$. The latter depends crucially on the Prandtl number: in the case of low-Prandtl fluids, a naive laminar estimate would lead to a thermal boundary layer thicker than the kinematic one. However, one expects vigorous turbulent mixing outside the kinematic boundary layer, so that both boundary layers end up having the same thickness. We obtain $\delta_\Theta \sim H / \sqrt{\reynolds}$.}
Upon estimating the Reynolds number using the free-fall velocity scaling \com{(based on $\deltaTBL$ in a worst-case scenario)}, equation~(\ref{eq:temperature_jump_BL}) becomes: 
\begin{equation}
\left( \deltaTBL  \right) ^3 \sim \frac {P^2H}{\alpha g \rho^2 \heatcap^2 \ell^2} \frac{\nu^2}{\kappa^2}\, . \label{eq:deltaT_stagnant_lowPr}
\end{equation}
which \com{remains} negligible compared with the bulk temperature jump from equation~(\ref{eq:nusselt_RaPr_bulk}) when $\prandtl \ll 1$. As a consequence, the scaling-law $\nusselt \sim \dimlessell \sqrt{\rayleigh \prandtl}$ remains valid and is not modified by the stagnant bottom layer for $\prandtl \ll 1$.


\cor{The situation is crucially different for high-Prandtl fluids. The} thermal boundary layer is much thinner than, and nested into, the kinetic one. The flow in the vicinity of the no-slip bottom surface is perceived by the thermal boundary layer as a parallel shear flow, the \cor{bottom shear $S(0)$ being estimated as} $S(0) \sim U / \delta_\nu \sim U \sqrt{\reynolds}/ H \sim \sqrt{{U^3}/{H\nu}}$.
The boundary-layer \cor{thickness is set by a balance between slow} advection by the shear\cor{, at a speed $S(0)\delta_\Theta$, and vertical diffusion: }
\begin{equation}
\rho \heatcap S(0) \delta_\Theta \frac{\deltaTBL}{H} \sim \frac{\rho \heatcap \kappa \deltaTBL}{\delta_\Theta^2}\,, \qquad \mathrm{so\: that} \qquad 
\frac{\delta_\Theta}{H} \sim\left(\frac{\kappa}{S(0)H^2}\right)^{1/3} \sim \frac{1}{\reynolds^{1/2}\prandtl^{1/3}}\,.
\label{eq:sheared_BL_noslip}
\end{equation}
Injecting this expression for the thermal boundary layer thickness into equation~(\ref{eq:temperature_jump_BL}), one gets the temperature jump associated with the sheared thermal layer and the associated dimensionless \cor{heat flux}:
\begin{equation}
\left( \deltaTBL  \right) ^3 \sim \frac {P^2H}{\alpha g \rho^2 \heatcap^2 \ell^2}\left(\frac{\nu}{\kappa}\right)^{2/3}\,;\quad \nusselt \sim \dimlessell \rayleigh^{1/2}\prandtl^{1/6}\, . \label{eq:deltaT_stagnant_highPr}
\end{equation}
Noticeably, a comparison with equation (\ref{eq:nusselt_RaPr_bulk}) informs us that the temperature jump in the stagnant boundary layer dominates the temperature jump in the bulk\com{: $\Delta T \simeq \deltaTBL$} \cor{when $Pr \gg 1$}. This observation results in a modification of the Nusselt number scaling-law due to boundary layer corrections\com{, see equation (\ref{eq:deltaT_stagnant_highPr})}.

The last case that requires our attention is the case of high-Prandtl-number flows with a stress-free bottom surface. Within the thermal boundary layer, the flow is well approximated by a uniform parallel flow with velocity $U\sim \sqrt{\alpha g \deltaTBL H}$\modif{. Here, in view of finding hypothetical boundary layer corrections,} \com{we assume again that the overall temperature difference is dominated by the contribution from the boundary layer $\deltaTBL$, before testing this assumption {\it a posteriori}.} A triple balance between horizontal advection, diffusion through the layer, and heating yields:
\begin{equation}
\rho \heatcap \sqrt{\alpha g \deltaTBL H} \frac{\deltaTBL}{H} \sim \rho \heatcap \kappa \frac{\deltaTBL}{\delta_\Theta^2} \sim \frac{P}{\ell} \,.
\label{eq:highPr_SF_balance}
\end{equation}
This balance is similar to the bulk balance (eq.~\ref{eq:bulk_balance}). Some straightforward algebra confirms that a transport law identical to (\ref{eq:nusselt_RaPr_bulk}) is found. We conclude that no boundary-layer induced corrections to the heat-transport \com{scaling-law} are necessary in this case.

\cor{To summarize, the scaling theory results in the following predictions for the dimensionless heat flux:}
\begin{equation}
\nusselt \cor{=}
\left\lbrace
\begin{array}{ccc}
\cor{c_1} \, \dimlessell \sqrt{\rayleigh \prandtl} & \mbox{with stress-free b. c., } & \forall \prandtl \, ,\\
\cor{c_2} \,\dimlessell \sqrt{\rayleigh \prandtl} & \mbox{with no-slip b. c., } & \prandtl \ll \prandtl_* \, , \\
\cor{c_3} \,\dimlessell \rayleigh^{1/2} \prandtl^{1/6} & \mbox{with no-slip b. c., } & \prandtl\gg \prandtl_* \, , 
\end{array}\right.
\label{eq:nusselt_sumup}
\end{equation}
\cor{where the dimensionless prefactors $c_i$ remain to be determined. The transition Prandtl number $\prandtl_*$ is estimated by equating the two no-slip predictions at $Pr=Pr_*$, which yields $Pr_*=(c_3/c_2)^3$.}

\begin{figure}
  \centerline{\includegraphics[height=0.5\textwidth]{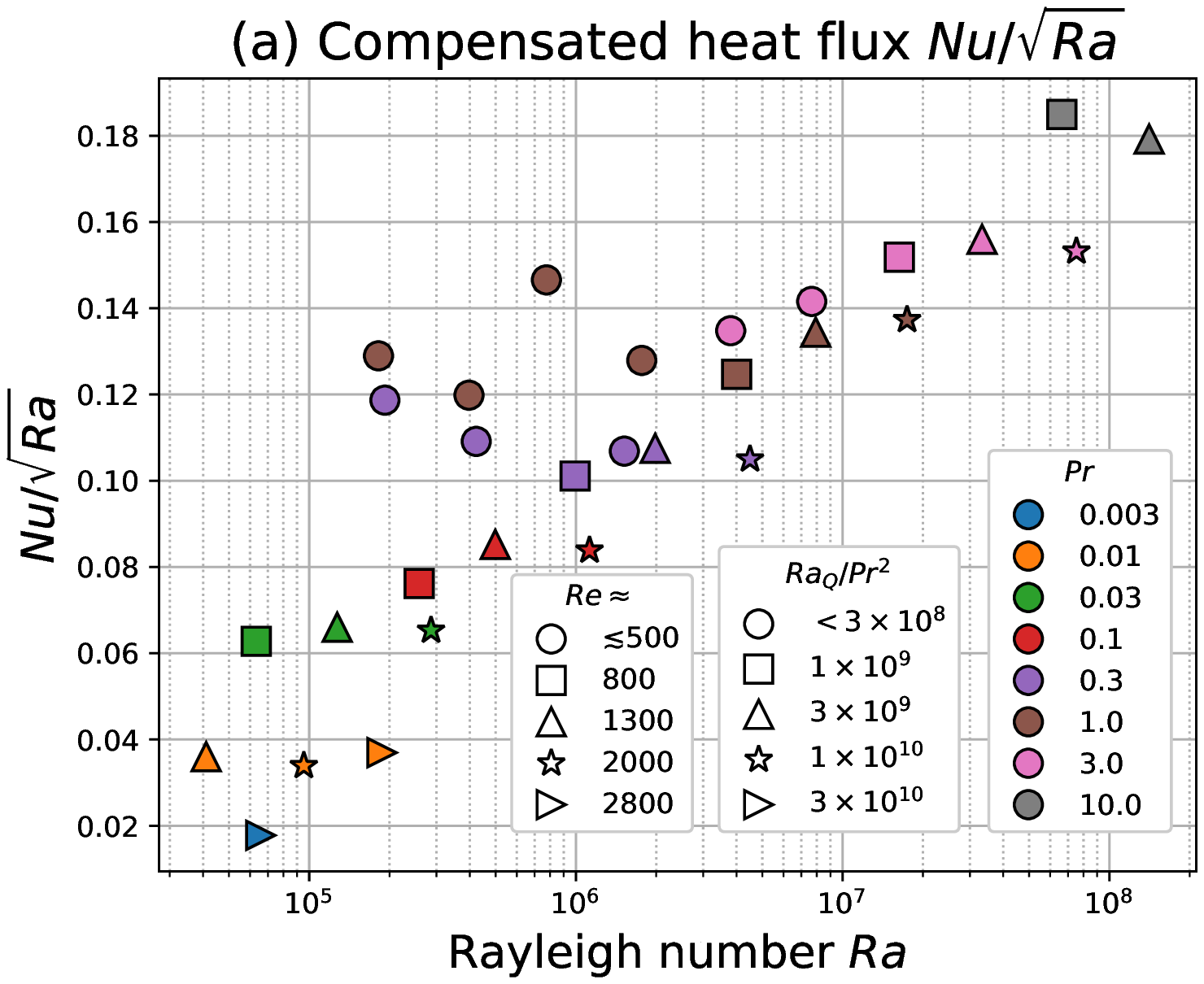}}
  \centerline{\includegraphics[height=0.5\textwidth]{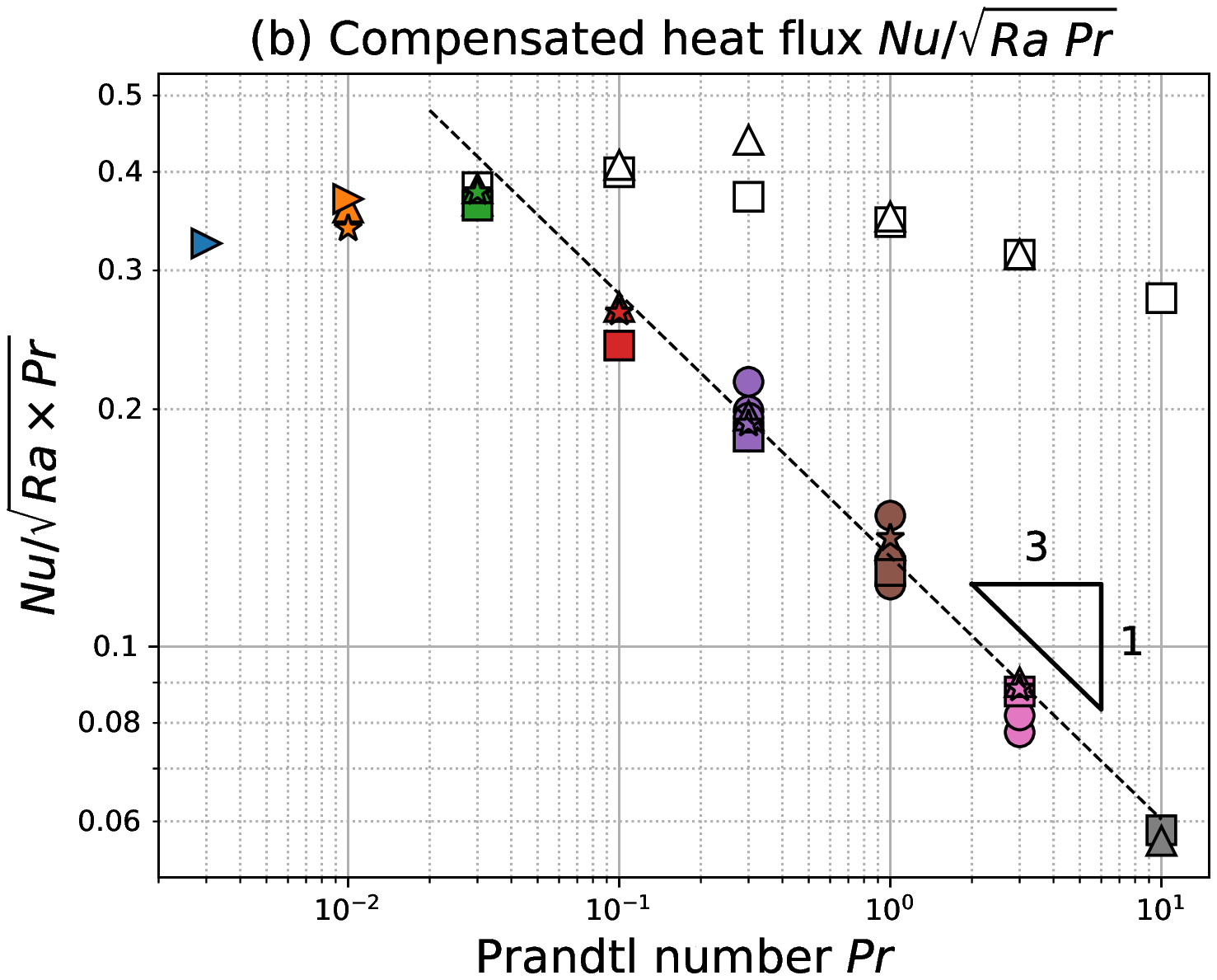}}
  \caption{(a) \cor{Compensated heat-flux $\nusselt / \sqrt{\rayleigh}$ as }a function of the Rayleigh number $\rayleigh$ for no-slip boundary conditions. \cor{The colour codes for} the Prandtl number $\prandtl$, while the heating intensity as measured by $\rayleighQ/\prandtl^2$ is encoded in the shape of the symbols. For convenience, we provide a correspondence between the parameter $\rayleighQ/\prandtl^2$ and the Reynolds number $\reynolds$. This colour/shape code is employed consistently thereafter throughout the paper. (b) Compensated heat-flux $\nusselt/ \sqrt{\rayleigh \prandtl}$ as a function of $\prandtl$ for no-slip (\cor{filled} symbols) and stress-free (open symbols) bottom boundaries.\label{fig:nu_f(Pr)}}
\end{figure}


\section{Observations of the two regimes}\label{sec:2regimes}

\cor{We now turn to an extensive numerical study, with the goal of (i) assessing the validity of the scaling predictions (\ref{eq:nusselt_sumup}), (ii) estimating the transition Prandtl number $\prandtl_*$, and (iii) determining the dimensionless prefactors $c_i$, so that (\ref{eq:nusselt_sumup}) can be used as a quantitative relation in cases of practical interest. Throughout this section, the dimensionless absorption length is set to $\dimlessell=0.1$. The domain is cubic with periodic boundary conditions in the horizontal and insulated top and bottom boundaries. The top boundary condition is always no-slip, and the bottom one is either stress-free or no-slip.}

\subsection{Heat flux scalings}
\cor{Guided by the} prediction~(\ref{eq:nusselt_sumup}) and the observations by~\citet{lepotPNAS18} for $\prandtl=1$ and $7$, we represent on figure~\ref{fig:nu_f(Pr)}a the Nusselt number $\nusselt$ compensated by $\sqrt{Ra}$ for data obtained in a suite of DNS runs with almost logarithmically-equispaced Prandtl number \cor{values}: $\prandtl \in \left\{0.003,\, 0.01,\, 0.03,\, 0.1,\, 0.3,\, 1,\,3,\,10\right\}$. We observe that for a fixed Prandtl number (a given colour on the figure), the data form a plateau \cor{as} the thermal forcing $\rayleighQ$ varies. This \cor{confirms} that the scaling exponent $\rayexpo=0.5$ is universal and does not depend on the Prandtl number\cor{, in line with} prediction~(\ref{eq:nusselt_sumup}).

We now turn to the central question of this paper -- the Prandtl number dependence -- which amounts to analysing the height of the plateaus on figure~\ref{fig:nu_f(Pr)}a. On this logarithmic representation, \cor{the} plateaus are separated by a distance that \cor{seems} to vary with the Prandtl number.
 
To better characterise this effect, we turn to figure~\ref{fig:nu_f(Pr)}b where we plot the heat flux compensated by $\sqrt{\rayleigh\prandtl}$ -- the \cor{scaling prediction}~(\ref{eq:nusselt_sumup}) in absence of boundary-layer \cor{correction} -- as a function of the Prandtl number. Upon this rescaling, \cor{the} data points for no-slip boundary conditions (\cor{filled} symbols) fall on a master curve that exhibits: (i) a plateau for $\prandtl\lesssim 0.04$, providing evidence for the $\nusselt\sim \sqrt{\rayleigh\prandtl}$ scaling at low Prandtl numbers; (ii) a scaling-law with slope $-1/3$ for $\prandtl \gtrsim 0.04$ which, when combined with the rescaling of the figure, suggests the $\nusselt \sim \rayleigh^{1/2}\prandtl^{1/6}$ scaling predicted at high Prandtl numbers.

\cor{By contrast,} the data points obtained for stress-free boundary conditions (open symbols) reasonably form a plateau, as they exhibit variations of less than $30\%$ \cor{when $\prandtl$ spans three and a half decades.} This observation validates the mixing-length scaling $\nusselt \sim \sqrt{\rayleigh \prandtl}$ at all Prandtl \cor{numbers} for stress-free boundary conditions.

\cor{To summarize, the numerical data are fully compatible with the prediction~(\ref{eq:nusselt_sumup}), with the following values for the dimensionless prefactors and transition Prandtl number: $c_1=c_2 \simeq 0.38$, $c_3\simeq 0.13$, which leads to $Pr_*=(c_3/c_2)^3 \simeq 0.04$.}

\begin{figure}
  \centerline{\includegraphics[height=0.5\textwidth]{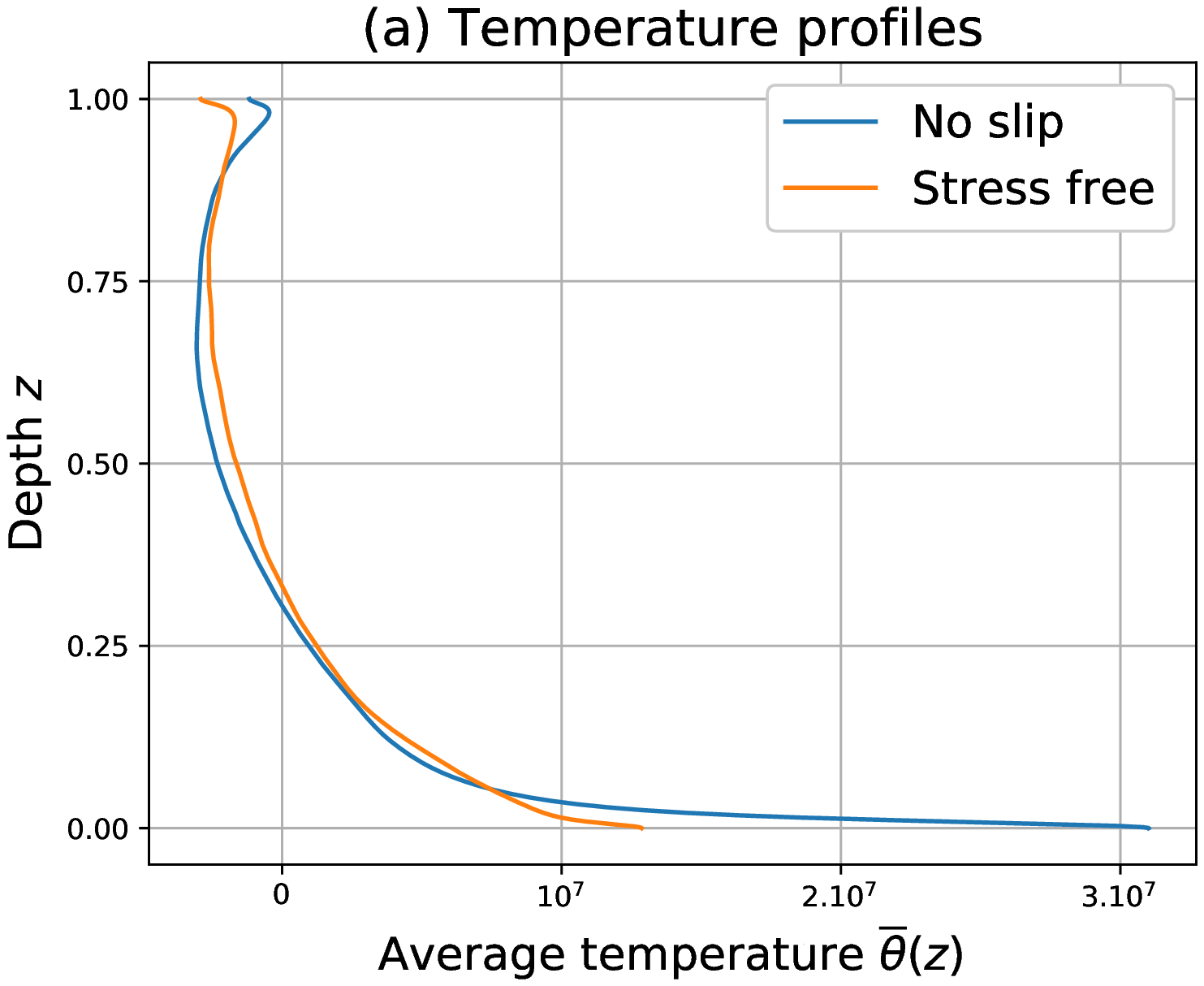}
              \includegraphics[height=0.5\textwidth]{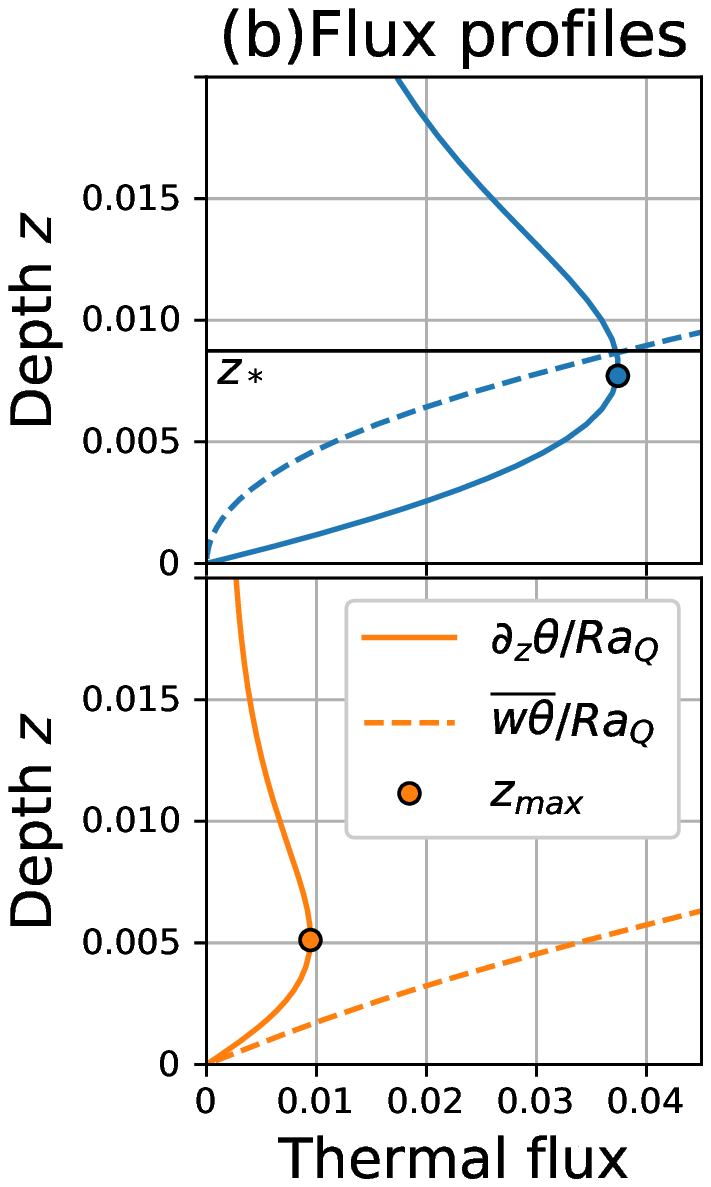}}
  \caption{(a) Temperature profiles averaged in time and along horizontal directions $x$ and $y$, for $\prandtl=3$, $\rayleighQ=3\times 10^{10}$, with either a no-slip (blue) or  a stress-free (orange) bottom boundary. (b) Time-and horizontally averaged contributions to the total heat-\modif{flux}. Solid lines represent the diffusive flux $-\overline{\partial_z \theta}$, whereas dashed lines represent the convective flux $\overline{w\theta}$. Coloured circles materialise $z_{\mathrm{max}}$, the altitude where the diffusive flux is maximum. For a no-slip bottom boundary (top panel), the thin horizontal black line represents the altitude $z_*$ where the diffusive and the convective fluxes are equal (see text).\label{fig:profiles} }
\end{figure}
\begin{figure}
	\centerline{\includegraphics[height=0.45\textwidth]{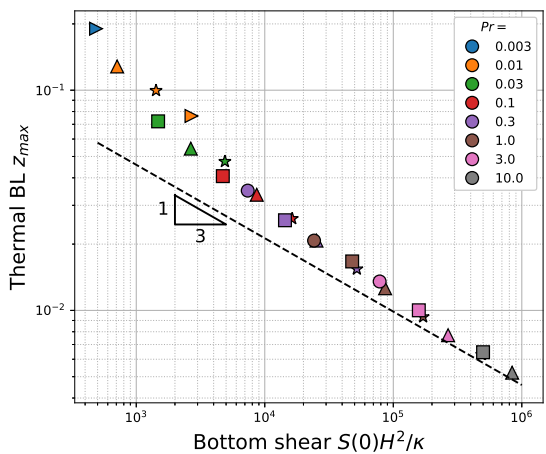}}
	\caption{Structure of the boundary layers as $\prandtl$ and $\rayleighQ$ vary. We represent the dimensionless thickness of the thermal boundary layer $\delta_\Theta/H$ defined as the depth where the diffusive flux is maximum (see figure~\ref{fig:profiles}b), represented as a function of the dimensionless shear at the bottom plate $S(0)H^2/\kappa =\overline{\partial_z\sqrt{u^2+v^2}}$. Dashed line: prediction from equation~(\ref{eq:sheared_BL_noslip}). \label{fig:BL_structure} }
\end{figure}

\subsection{Boundary layer structure: no-slip case}

We have established that, in the \com{high-Prandtl-number} regime $\prandtl\gtrsim 0.04$, thermal transfer laws differ for \cor{a no-slip and a stress-free bottom boundary and are accurately captured by the scaling prediction~(\ref{eq:nusselt_sumup}). To validate \emph{a posteriori} the underlying assumptions of the model, we now examine the structure of the boundary layers in the DNS flows.}

The structure of the thermal boundary layer is illustrated \cor{in} figure~\ref{fig:profiles} for a value of the prandtl number $\prandtl=3$ that lies comfortably within the high-Prandtl-number regime, where a clear departure is observed between no-slip and stress-free bottom boundaries. On this figure, we display the vertical temperature profiles $\overline{\theta}$, where the overline denotes an average along the horizontal directions and over time.
Upon examination, the temperature profiles are strikingly similar for no-slip and stress-free in the bulk of the flow, while \cor{the no-slip case develops} strong boundary layers associated \cor{with} a large temperature jump at the bottom of the layer. 

Beyond the mere temperature profiles, we analyse the thermal energy fluxes in the vicinity of the bottom bounding surface on figure~\ref{fig:profiles}(b). For no-slip boundary conditions, the diffusive flux $\overline{\partial_z \theta}(z)$ grows linearly with the distance from the wall $z$. As a consequence, the diffusive flux strongly dominates the purely convective flux $\overline{w\theta}(z)$, which exhibits a slower quadratic growth with $z$. As $z$ increases, the diffusive flux reaches a maximum at distance $z_\mathrm{max}$ and decreases again. \cor{Advection and diffusion contribute in equal proportions to the heat transfer at a comparable height $\zequi$, where $\overline{\partial_z \theta} (\zequi) = \overline{w\theta}(\zequi)$. Above $\zequi$ is the bulk, where pure convection proves the most efficient mechanism.}

Being able to extract a thermal boundary layer thickness from \cor{the} numerical solutions, we now turn our attention to the validity of \cor{the} sheared boundary layer asumption~(\ref{eq:sheared_BL_noslip}) \cor{near a no-slip boundary}. \cor{The dimensionless bottom shear $S(0)H^2/\kappa$ is estimated as}:
\begin{equation}
	{S(0)H^2}/{\kappa} = \overline{\partial_z\sqrt{u^2(x,y,z,t) + v^2(x,y,z,t)}}|_{z=0}
	\,.
\label{def:shear}
\end{equation}
The boundary layer thickness measured by $z_\mathrm{max}$ is displayed on figure~\ref{fig:BL_structure} as a function of \cor{this dimensionless bottom shear}, for all the data gathered with no-slip boundary conditions. A good collapse is observed, and \cor{the data for large Prandtl number validate} the shear layer structure assumed in equation~(\ref{eq:sheared_BL_noslip}). \com{The approach to this asymptote proves rather slow, because the kinetic boundary layer is only moderately thicker than the thermal one for the highest $Pr$ achieved in this study (approximately $3$ times thicker for $Pr=10$).}

\subsection{Boundary layer structure: stress-free case}
Returning to figure~\ref{fig:profiles}(b), the examination of the energy flux profiles in the stress-free case yields a much different scenario. In contrast with the no-slip case, both the convective flux $\overline{w\theta}$ and the diffusive flux $\overline{\partial_z \theta}$ grow linearly \cor{with} the distance $z$ to the \cor{bottom boundary}. As a first consequence, the ratio between these two fluxes remains asymptotically of order $O(1)$ as the distance to the wall decreases. Moreover, energy transfers are dominated by pure convection as $\overline{w\theta}>\overline{\partial_z\theta}$ for any altitude $z$. All these observations are in strong support of the working hypotheses used for deriving prediction~(\ref{eq:nusselt_sumup}), and thus shed light on the (non-purely-diffusive) nature of the boundary layers that result in the $\nusselt \sim \sqrt{\rayleigh \prandtl}$ scaling observed for all $\prandtl$ on figure~\ref{fig:nu_f(Pr)}. 

\section{Discussion and conclusion}\label{sec:discussion}



We have characterised the dimensionless heat flux arising in convection driven by internal heat sources and sinks, focusing on the influence of the diffusivity ratio. \com{Combining the present results with the conclusions of \citet{lepotPNAS18} and \citet{bouillautJFM19}, we have answered the question raised at the outset -- the scaling behaviour of the enhanced heat flux -- by proposing the compact scaling relation (\ref{eq:nusselt_sumup}) for the  Nusselt number. While the scaling exponents appearing in this expression are most likely robust features of CISS, the precise values of the prefactors $c_1$, $c_2$ and $c_3$ may very well depend on details of a given system. For instance, extracting the prefactor $c_3$ from the experimental data obtained by \citet{lepotPNAS18} in a cylindrical container with a stress-free top boundary leads to approximately $0.9$. Their DNS data in a cube with stress-free sidewalls yield a value closer to $c_3\simeq1.1$, while the present DNS data with periodic boundary conditions in the horizontal lead to $c_3\simeq1.3$.}

When a stress-free bottom boundary is considered, we have established that the heat flux follows the standard mixing-length prediction $\nusselt \sim \sqrt{\rayleigh\prandtl}$, both in terms of Rayleigh and Prandtl number. 
When the bottom boundary condition is no-slip, the mixing-length scaling regime is observed at low Prandtl number, $Pr \leq 0.04$. As most astrophysical fluids of interest have a low-Prandtl number, this confirms that the associated flows would lead to the mixing-length scaling regime regardless of the bottom boundary condition. The present setup thus offers a clear realisation of the scaling-laws believed to hold in astrophysical contexts.

However, for $\prandtl>0.04$ and a no-slip bottom boundary, we find that the stagnant layer in the vicinity of the latter is conducive to a modification of the Prandtl-number exponent: $\nusselt\sim \rayleigh^{1/2}\prandtl^{1/6}$. This scaling is well captured by modelling the bottom thermal boundary layer as a sheared layer nested inside the \com{kinetic} boundary layer. We confirmed this picture through an analysis of the temperature and heat flux profiles extracted from our suite of DNS. This $Pr={\cal O}(1)$ situation could be relevant to the atmosphere, in situations where the latter is subject to both infrared heating and cooling~\citep{deardorffBLM74}. The no-slip bottom boundary and the associated stagnant layer would then crucially impact the temperature profile in the atmospheric boundary layer. But this situation is also relevant to laboratory realisations of CISS~\citep{lepotPNAS18,bouillautJFM19}, which use water ($Pr=7$) together with a no-slip bottom boundary. Several approaches can be envisioned to achieve the $\nusselt \sim \sqrt{\rayleigh\prandtl}$ \modif{scaling} in such a laboratory experiment. Following the numerical study of~\citet{barkerApJ14}, one approach would consist in artificially picking a distribution of sources and sinks that vanishes at the top and bottom boundaries, but there is no simple experimental way of tailoring the heat source distribution. A more natural approach from the experimental standpoint would be \com{to} use a fluid with a Prandtl number much below $Pr_*=0.04$, but that again seems rather impracticable. Finally, our study suggests a third and much more promising route towards observing the $Pr^{1/2}$ dependence of the Nusselt number experimentally: tweaking the bottom boundary condition to make it as close as possible to a stress-free one, by stacking two layers of immiscible fluids, a heavy transparent one below a lighter light-absorbing one.

\modif{This work was performed using HPC resources from GENCI-CINES (Grant 2019-A0060410803 and Grant 2020-A0082A10803). }\com{This work is supported by the European Research Council under grant agreement 757239.}

Declaration of Interests. The authors report no conflict of interest.

\bibliographystyle{jfm}
\bibliography{prandtl_convection}

\end{document}